\documentclass[5p]{elsarticle}

\usepackage{graphicx}
\usepackage{dcolumn}
\usepackage{bm}
\usepackage{longtable}
\usepackage{float}
\usepackage[fleqn]{amsmath}

\makeatletter
\def\ps@pprintTitle{%
 \let\@oddhead\@empty
 \let\@evenhead\@empty
 \def\@oddfoot{}%
 \let\@evenfoot\@oddfoot}
\makeatother

\usepackage{lineno,hyperref}
\modulolinenumbers[5]

\begin{document}

\begin{frontmatter}

\title{Demagnetizing fields in all-optical switching}

\author{F. Hoveyda}

\author{E. Hohenstein}

\author{R. Judge}

\author{S. Smadici \corref{cor1}}


\address{Department of Physics and Astronomy, University of Louisville, KY 40292, USA}%

\begin{abstract}
Time-resolved pump-probe measurements show ultrafast and heat accumulation demagnetization in Co/Pd superlattices on glass substrates. A model of demagnetizing fields and micromagnetic simulations are applied to examine the evolution of a demagnetized cylinder into a switched state.
\end{abstract}


\end{frontmatter}


\section{Introduction}

Ultrafast demagnetization (UDM)~\cite{1996Beaurepaire} is not an ultrafast rotation of magnetization. Instead, the magnetization partially or completely fragments in the strong pulsed laser field over a demagnetization time $\tau_{M}$. A large exchange energy is obtained and re-assembly in the initial direction occurs over an equilibration time $\tau_{E}$. Early links between $\tau_{M,E}$ and spin precession damping rates~\cite{2005Koopmans} were developed into quantum models with spin-flip electron scattering in transition metal or rare earth materials and alloys~\cite{2010Koopmans,2014Kuiper,2014Gunther,2014Mendil} (figure 1(a), left panel). Dependence on external magnetic field~\cite{2016Tsema}, ambient temperature~\cite{2010Koopmans,2012Roth}, and excitation wavelength~\cite{2017Bierbrauer,2017Bobowski} have been examined. Alternative models~\cite{2008Kazantseva,2015Hinzke} and transient spin currents~\cite{2010Battiato,2013Schellekens-a,2014Schmising,2015Schmising,2017Eschenlohr} have been considered. Experiments on heterostructures, with electron diffusion from a heating Al layer into Ni~\cite{2013Eschenlohr,2016Salvatella}, in Co/Pd multilayers~\cite{2016Vodungbo}, and with ballistic electrons through a Cu layer into Co/Pt~\cite{2016Bergeard}, showed that it is possible to obtain UDM without direct light interactions.

In contrast, all-optical switching (AOS) is the light-induced reversal of magnetization to a single-domain state in ferrimagnetic films~\cite{2007Stanciu}, ferromagnetic films~\cite{2014Mangin,2014Lambert,2016Hadri-a,2016Hadri-b,2016Hadri-c,2017F}, and granular media~\cite{2014Lambert,2016Takahashi,2017John} with perpendicular magnetic anisotropy. A complete UDM persists at the center of the excitation area with no discernible equilibration time $\tau_{E}$ before a gradual emergence of AOS (figure 1(a), right panel) in GdFeCo~\cite{2014Hashimoto}. The transient magnetic field of the Inverse Faraday Effect, induced by the laser pulse in the material and harder to estimate in metals~\cite{2016Gorchon-a,2016Cornelissen} than in doped glass~\cite{1965Ziel}, has been considered as a possible explanation~\cite{2007Stanciu,2016Cornelissen,2016Nieves,2016Berritta,2010Kirilyuk}. Alternative models include the different absorption of magnetic circular dichroism~\cite{2016Gorchon-a,2012Khorsand,2016Ellis}, thermally induced magnetisation switching of ferrimagnets with linearly-polarized light~\cite{2012Ostler,2013Schellekens-b,2014Oniciuc,2015Atxitia}, atomic models~\cite{2017Murakami}, or models specific to granular media~\cite{2016Gorchon-a,2016Ellis}.

In this work, time-resolved pump-probe measurements were made on Co/Pd superlattices that show partial to complete demagnetization from ultrafast processes and heat accumulation. An analytical model and micromagnetic simulations are applied to determine the conditions for demagnetizing fields to nucleate and develop an AOS state from a demagnetized disk.

\section{Experiments}

\subsection{Setup}

\begin{figure}
\centering\rotatebox{0}{\includegraphics[scale=0.35]{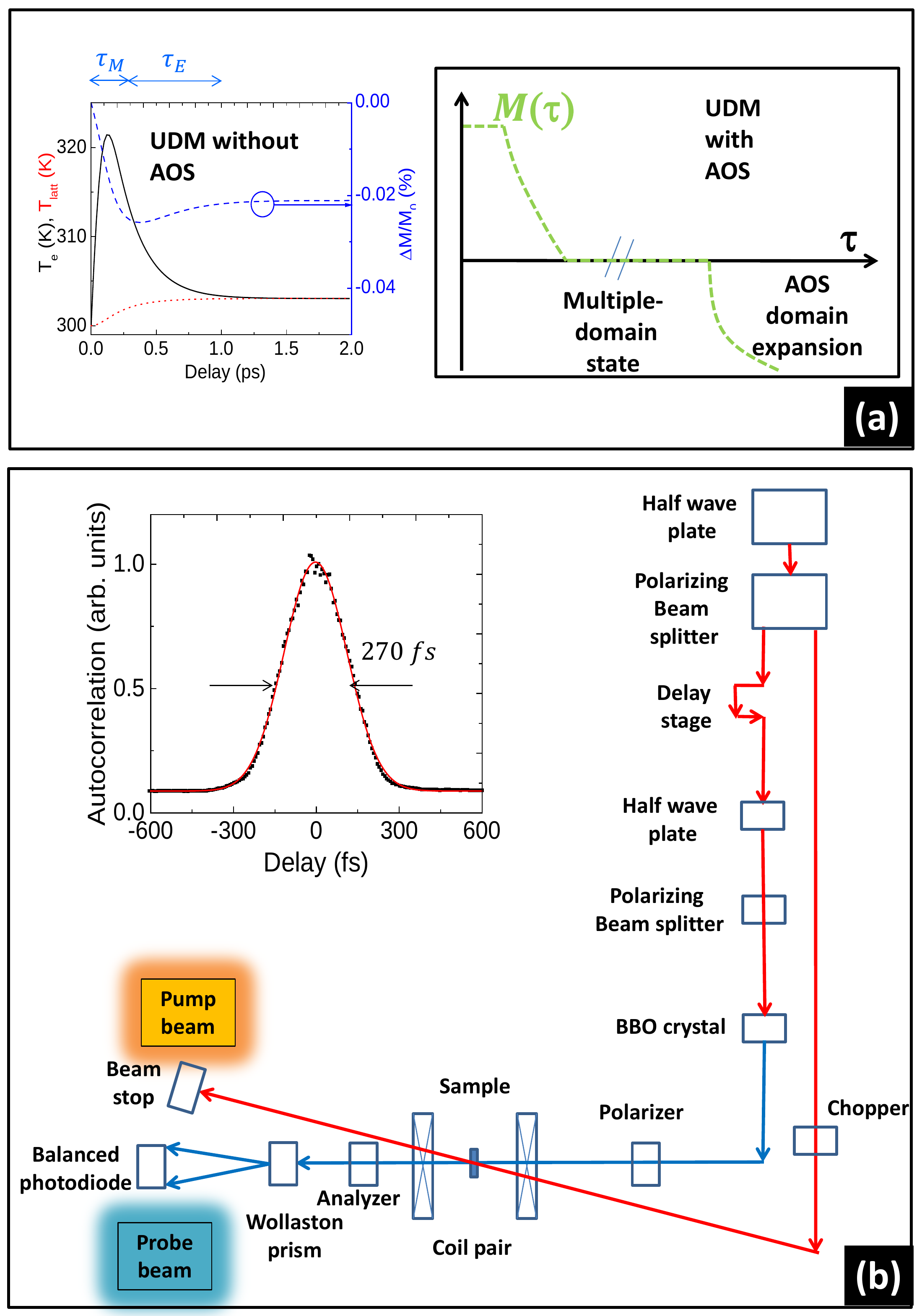}}
\caption{\label{fig:Figure1} (a) Left panel: $T_{e}, T_{latt}, M$ of a ultrafast model for $P_{abs}=40~mW$ and $h=4.1~nm$ with bulk cobalt parameters~\cite{2010Koopmans,2014Kuiper,2015Schmising}, neglecting heat accumulation and electron-induced UDM. Right panel: sketch of magnetization time-dependence in GdFeCo with AOS domains starting after $\sim 10~ps$ at the center~\cite{2014Hashimoto}. This takes a $ms$-range time interval in cumulative AOS of Co/Pd~\cite{2017F} and Co/Pt~\cite{2016Hadri-b}. (b) Sketch of the experimental setup. Inset: autocorrelation of 190 fs pump pulses at the sample location.}
\end{figure}

Ferromagnetic $\rm [Co/Pd]_{4}$ superlattices were examined, in which cumulative AOS was observed with linearly-polarized light~\cite{2017F}. The samples were $h=4.1~nm$ thick with perpendicular magnetic anisotropy (PMA).

The pump-probe two-frequency setup has a non-collinear geometry, with measurements in transmission at normal incidence (figure 1(b)). The linearly-polarized 800 nm pump and 400 nm probe beams were focused to stationary $w_{0}=125~\mu m$ and $w_{1}=80~\mu m$ spots, respectively, and the delay between the two pulse sequences scanned with a translation stage.

The sample magnetization was re-initialized between pump pulses with a constant field $|B|=300~G$ from two water-cooled coils. The relatively strong damping $\alpha\approx 0.1$ in Co/Pd~\cite{2011Liu,2011Pal} insures that the magnetization is stable within the $12.5~ns$ time interval between pulses. This allows measuring transient processes with the same initial and final states. Measuring the AOS time dependence, with different initial and final states, requires a field pulsed at the TiS laser repetition rate ($\rm 80~MHz$) and cannot be currently done in our setup.

A balanced photodiode lock-in detection at the pump beam chopping frequency $f=2.069~kHz$ has been applied. Intensity and polarization variations arise from temperature and birefringence transients. A configuration with a polarizer and analyzer near crossing minimizes non-magnetic contributions to the anti-symmetric component $A(\tau)$~\cite{2003Koopmans-book}. Measurements away from crossing configuration resulted in a featureless $A(\tau)$.

\subsection{Results}

\begin{figure}
\centering\rotatebox{0}{\includegraphics[scale=0.4]{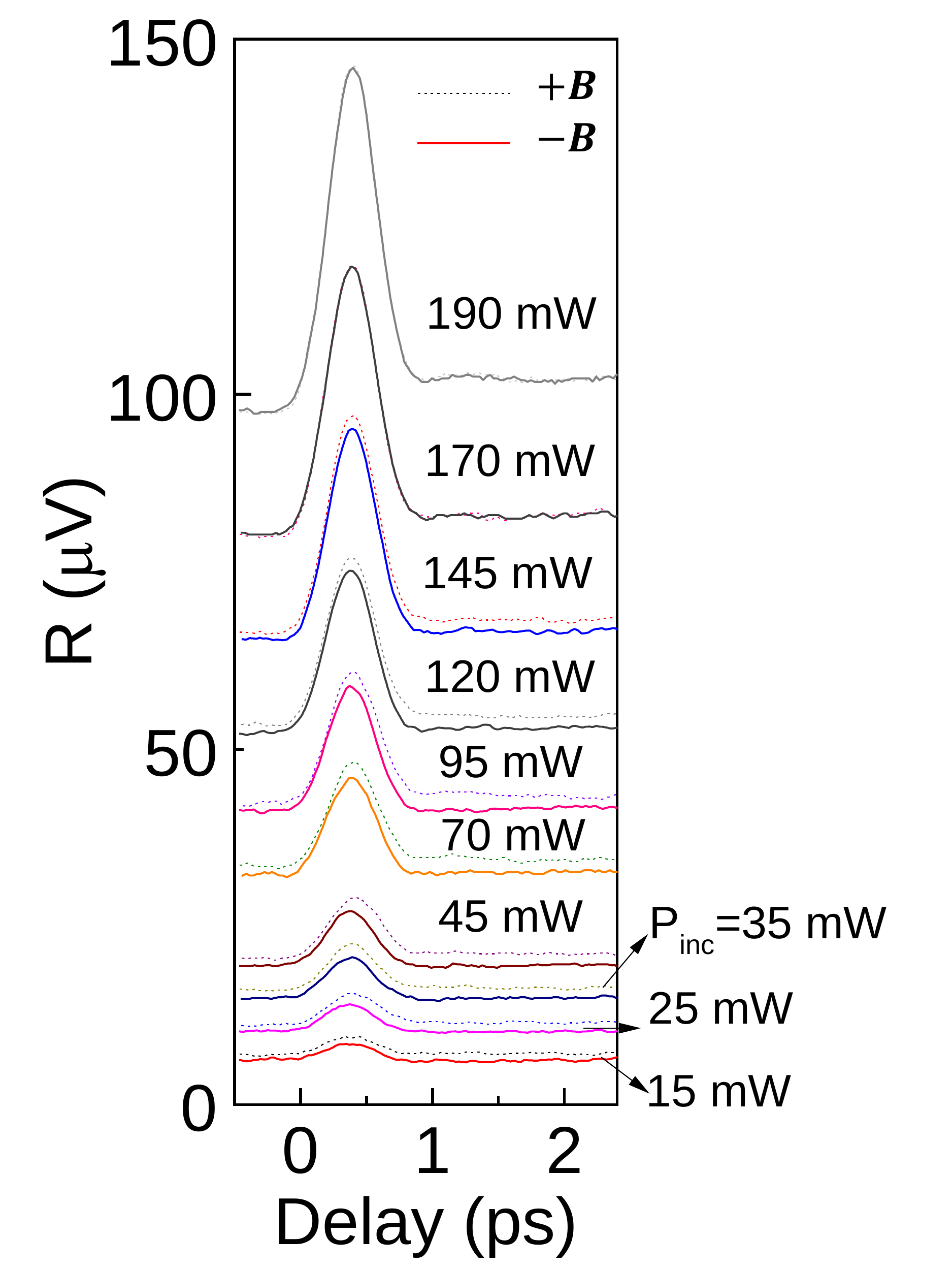}}
\caption{\label{fig:Figure1} $R(\tau)$ for different magnetic fields and incident pump beam power.}
\end{figure}

The lock-in resultant $R$ dependence on delay $\tau$ for different pump beam powers and applied fields $B=\pm 300~\rm G$ (figure 2) was separated into two components, $S_{R}(\tau)=\frac{1}{2}\Big( R(-B)+R(+B)\Big)$ and $A_{R}(\tau)=\frac{1}{2}\Big( R(-B)-R(B)\Big)$, symmetric and anti-symmetric in $B$, respectively.

The symmetric part $S(\tau)$ (figure 3) shows a prominent peak, similar to results for Co films~\cite{2005Bigot}, followed by a step. A glass substrate transient plasma~\cite{2007Siegel} contributes in our transmission geometry at relatively low fluence because of strong metal absorption. An example of electron and lattice temperature dependence in two-temperature models of energy exchange between electron and lattice thermal reservoirs~\cite{2000Hohlfeld} is shown in figure 1(a) for bulk cobalt parameters. Since glass also contributes and heat accumulation is large, a simpler Gaussian with a step fit has been applied, which describes the experimental results well (figure 3).

The film temperatures can be estimated for $E_{pulse,abs}=0.5~nJ$, corresponding to an absorbed power $P_{abs}=40~mW$. From the energy conservation, $T_{e,max}\approx \sqrt{\frac{2E_{pulse,abs}}{\gamma hw_{0}^{2}}+T_{0}^{2}}\approx 335~K$, when neglecting the transfer of energy to the lattice over the duration of the pulse, where $T_{0}=300~K$ is the initial temperature, $C_{e}=\gamma T$ with $\gamma=665~J/(m^{3}K^{2})$ for bulk Co~\cite{2014Kuiper}. Similarly, the lattice temperature step increase after one pulse is $T_{latt}\approx E_{pulse}/(C_{latt} hw_{0}^{2}) \approx 2.5~K$. In addition, the heat accumulation temperature $T_{acc}$ cannot be neglected for thin samples and high-repetition rate lasers, with multiple pulses incident on the same area within the heat diffusion time $\frac{w_{0}^{2}}{4D}$. The maximum heat accumulation is $T_{acc,max}=310~K$ for $P_{abs}=40~mW, h=4.1~nm, w_{0}=125~\mu m$, an interface conductance $G>10^{6}~W/m^{2}K$ and the same thermal parameters as in Ref.~\cite{2017F-arXiv}.

The antisymmetric part $A(\tau)$ (figure 4) increases and then decreases with power. The time-dependence corresponds to type I UDM~\cite{2010Koopmans}, with a small demagnetization time $\tau_{M}$, similar to Co and Co/Pt~\cite{2014Kuiper}, and consistent with measurements of UDM in Co/Pd with XMCD~\cite{2010Boeglin}, XRMS~\cite{2016Vodungbo}, and X-ray Fourier transform holography~\cite{2014Schmising}. In contrast to previous measurements, heat accumulation temperature $T_{acc}$ is significant and the UDM depth decreases at higher power, as magnetization is gradually removed. Complete demagnetization is obtained at $P_{inc}=170~mW$ which, from measurements of reflected and transmitted beam powers, corresponds to $P_{abs}=40~mW$. The heat accumulation temperature for this absorbed power gives a Curie temperature $T_{C}=610~K$, consistent with results in similar samples of $T_{C}=800~K$ for Co/Pd~\cite{2016Chen} and $T_{C}=600~K$ for Co/Pt~\cite{2014Kuiper}.

\begin{figure}
\centering\rotatebox{0}{\includegraphics[scale=0.4]{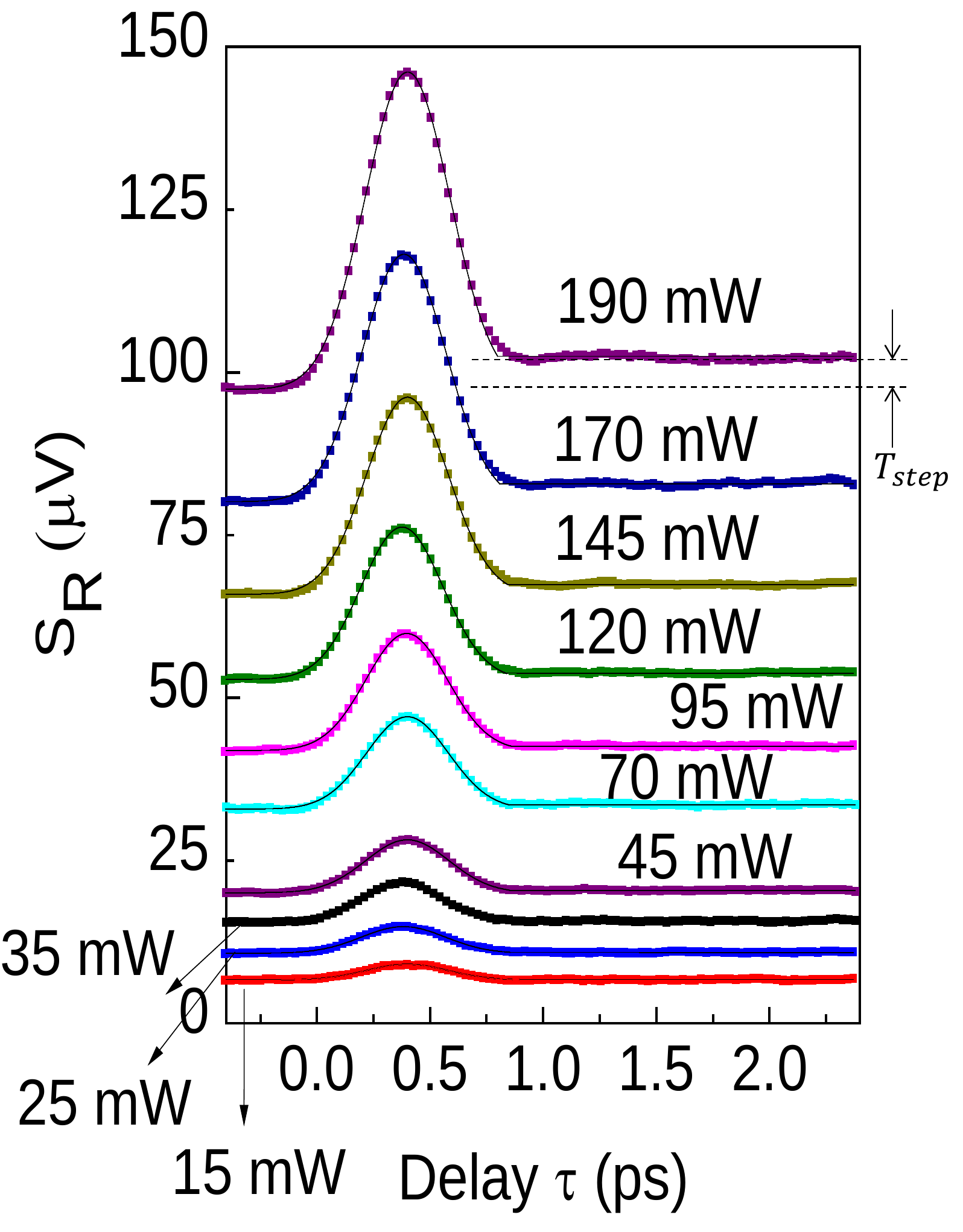}}
\caption{\label{fig:Figure1} $S(\tau)$ at different power and Gaussian with step fit.}
\end{figure}

\begin{figure}
\centering\rotatebox{0}{\includegraphics[scale=0.45]{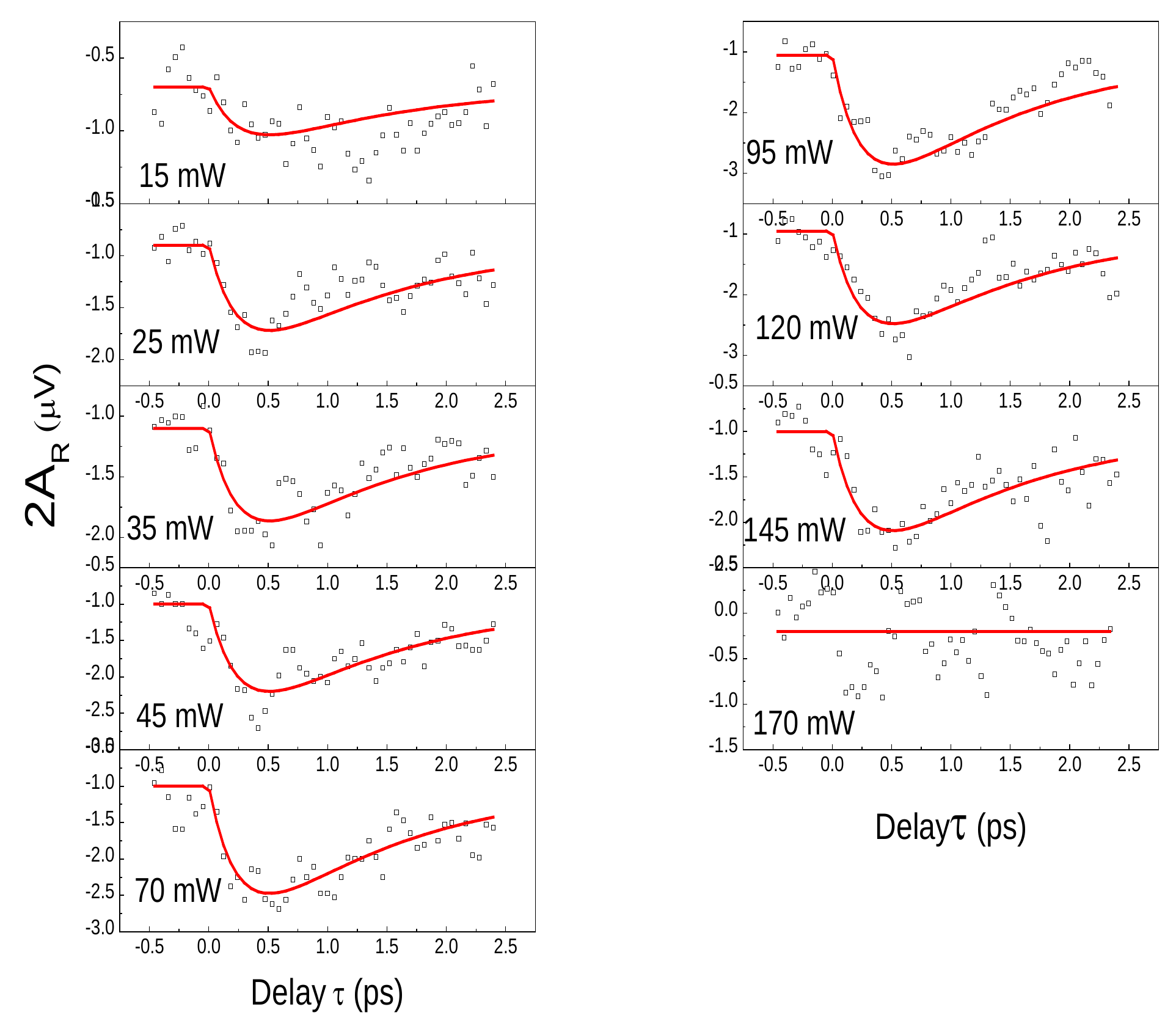}}
\caption{\label{fig:Figure1} $A(\tau)$ at different power and two-exponential fit.}
\end{figure}

The ultrafast time-evolution of magnetization in magnetic materials depends on the electron $T_{e}$ and lattice $T_{latt}$ temperatures. Transfer of energy from electrons to lattice includes a spin-flip process in a microscopic scattering model, with the magnetization variation $dM(\tau)/d\tau$ proportional to the rate of spin-flip scattering $R$~\cite{2010Koopmans,2014Gunther,2014Kuiper,2016Cornelissen}. This model requires several other parameters in addition to the Curie temperature $T_{C}$ and gives very good fits to the experimental results for transition metal and rare earth materials, with small and large $\tau_{M}$, respectively~\cite{2010Koopmans}. An example with bulk Co parameters is shown in figure 1(a). Indirect electron-induced UDM from the glass plasma~\cite{2007Siegel} may also contribute, as also seen in Co/Pd multilayers capped with Al~\cite{2016Vodungbo}. For simplicity, a rate equation fit was applied

\begin{eqnarray}
M(t)= A(P,T_{acc}) \Big( e^{-\frac{t}{\tau_{E}}}- e^{-\frac{t}{\tau_{M}}}\Big)\theta(t)+const.
\end{eqnarray}

\noindent where $\theta(t)$ is the step-function, and $\tau_{M,E}$ the demagnetization and equilibration times, respectively. The measurements were well fit with $\tau_{M}=0.25~ps$ and $\tau_{E}= 1.3~ps$. They confirm that demagnetized states are induced in Co/Pd. As in single-pulse AOS of GdFeCo~\cite{2014Hashimoto} and cumulative AOS of Co/Pt superlattices~\cite{2016Hadri-b}, a demagnetized state is a precursor of cumulative AOS in Co/Pd~\cite{2017F}. The relation between demagnetization and AOS was examined in a model of demagnetizing fields and with micromagnetic simulations.

\section{Discussion}

\subsection{Micromagnetic simulations}

Demagnetizing (DM) fields express the long-range magnetic moment dipolar interactions, which are relatively weak compared to the short-range exchange interactions. Nevertheless, DM fields and interactions cannot always be neglected, in particular in PMA materials. In practice, they can induce a precession of an intact macrospin~\cite{1999Back,2000Bauer}, have been proposed as a driver of AOS~\cite{1987Shieh}, and are relevant to AOS in TbFeCo~\cite{2009Ogasawara,2015Gierster}, in GdFeCo~\cite{2015Guyader,2016Guyader}, in ferromagnetic Co/Pt~\cite{2014Mangin,2014Lambert} and Co/Pd~\cite{2017F} superlattices.

A DM field $H_{D}$ will not switch the magnetization spontaneously in a PMA material because the anisotropy field $H_{K}=\frac{2K}{M_{s}}>4\pi M_{s} = H_{D}$. However, the initial state after the pump pulse is a demagnetized state, not a spatially-uniform PMA state. Investigating the time evolution of this state requires micromagnetic simulations.

Micromagnetic OOMMF~\cite{OOMMF-webpage} simulations were made, in which the classical macrospin was fragmented into cell spins with the cell size $a^3=(5~nm)^{3}$. The cell size sets the minimum feature size and can be compared to the Bloch domain wall (DW) width $l_{B}=\sqrt{\frac{A}{K}} \approx 6~nm$ for magnetic parameters at the center of the diagram and the width $l_{B}=25~ nm$ calculated for bulk Co~\cite{2016Moreno}. A demagnetized state is made in a cylinder with a radius comparable to beam size ($R=3-150~\mu m$) in AOS experiments. These volumes were impractical to simulate. Simulations of smaller volumes were made and the results compared to an analytical model. An initial random spin state was defined in a cylindrical volume of radius $R$ and height $h$ at the center of a $l\times l \times h$ plane (figure 5(a)). Typical values were $R=50~nm$, $h=10~nm$, and $l=1000~nm$. Runs with different $a,R$ and $h$ (not shown) gave results consistent with expectations.

The DM cylinder is a high-energy state because of spin misalignments. The time-evolution was examined with an energy minimization solver and the final state identified from its total magnetization normal to the plane $M_{z}$. A time-evolution solver, based on the Landau-Lifshitz-Gilbert equation with a damping constant $\alpha=0.05$, gave similar results.

The time-evolution is determined by macroscopic energy densities $K, A, 2\pi M^{2}_{s}$, where $K$ is an uniaxial anisotropy energy density with an easy axis normal to the surface, $A$ is the exchange stiffness $A=JS^{2}/d$, where $J$ is the nearest-neighbor exchange interaction and $d$ the structural unit cell, and $2\pi M_{s}^{2}$ is the DM field energy density. The uniaxial anisotropy $K$ tends to pull spins out of plane, the exchange interaction $A/a^{2}$ tends to keep spins aligned, and the DM field energy $2\pi M_{s}^{2}$ pulls them in plane. The process was automated and $K, A, M_{s}$ varied over a wide range. A complex phase diagram may be expected when $K$, $\frac{A}{a^{2}}$ and $2\pi M_{s}^{2}$ are the same order of magnitude and competing in determining the final magnetic state.

The magnetic structure of the initial state evolves into four different final states of lower energy: (1) a pattern made of small stable clusters, with spins pointing up or down (multiple domain, $MD$ state), (2) an expanding reversed domain ($S$), (3) a state with spins rotated to an in-plane direction ($IP$), and (4) a uniform ``no-change" state, when the domain closes ($NC$) (figure 5(a)). The dependence of the final state on $K$ and $A/a^{2}$ is shown for $M_{s}=1500\times 10^{3}~A/m$ and $M_{s}=500\times 10^{3}~A/m$, with each outcome of a run represented by the symbol at the respective position on the diagram (figure 5(b-c)). A sequence of images (supplementary figure 1) and a movie (supplementary material) show the time-evolution from the initial to the final state. Coalescence of the randomly-oriented spins into a reversed domain $S$ state and subsequent domain expansion is obtained over a range of $A/a^{2}$ and $K$. The $S$ state is not obtained when the DM fields are neglected (not shown).

\begin{figure}
\centering\rotatebox{0}{\includegraphics[scale=0.3]{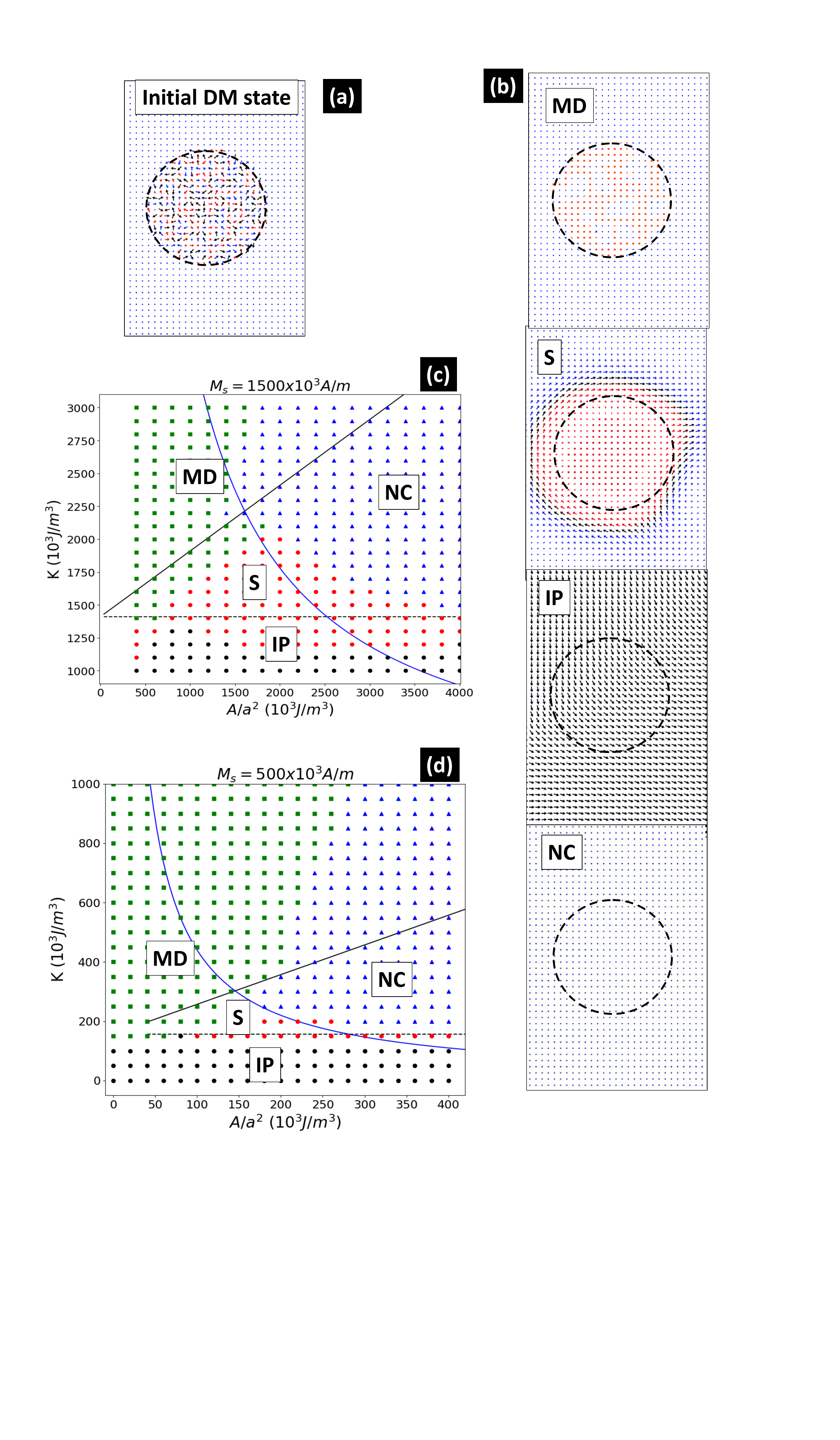}}
\caption{\label{fig:Figure4} (a) Initial random spin distribution in a $50~nm$ radius cylinder. Spins pointing up or down out of plane are labelled blue and red. (b) Examples of a $MD$ final state, $S$ state as domains walls expand outward beyond the radius of the initial cylinder, $IP$ state with spins in-plane, and $NC$ state with all spins pointing up. (c) Simulation results for $M_{s}=1500\times 10^{3}~A/m$, $h=10~nm$, and $R=50~nm$. (d) For $M_{s}=500\times 10^{3}~A/m$, $h=10~nm$, and $R=50~nm$.}
\end{figure}

The nucleation and expansion of the $S$ state in micromagnetic simulations can be understood in a continuum model. The horizontal line in figure 5(b-c) is the PMA boundary condition $K=2\pi M_{s}^{2}$, below which the magnetic state has in-plane spins. A multiple domain $MD$ state occurs when $K$ is large and the two minima for the two spin orientations along the easy axis are separated by a large barrier. Unless $A$ is also large, deepening one minimum over the other, depending on the orientation of neighbouring spins, the barrier cannot be overcome. The boundary between the $S$ and $MD$ states has a positive slope, with progressively larger $A$ required at larger $K$ to avoid the $MD$ state.

The parabola is the condition $p_{w}=p_{D}$, where $p_{w} \approx \frac{4\sqrt{AK}}{R}$ is the DW pressure that points inward, due to a reduction of DW energy with a smaller radius. $p_{D}$ is the DM field pressure. Pressures or forces per unit DW area represent the difference in energies on the two sides of a DW~\cite{1971Thiele}. Specifically, if $\Delta E$ is the difference in the energy density on the two sides of a DW, its displacement over a distance $d$ will change the energy by $(Ad)\Delta E$. This change is equal to the work done by a force $W=Fd=(Ap)d$ and the pressure on the DW is equal to the difference in the energy density on its two sides $p=\Delta E$. DM fields and pressures have been calculated before for a reversed cylinder in a macroscopic model in analogy to the electric fields of charged layers, with electric charges replaced by magnetic poles and integrating over poles distributed on the two surfaces $H_{D}(z)=\int \frac{M_{s} cos \alpha dA}{r^{2}}$, where $\alpha$ is the angle from the surface normal (figure 6(a))~\cite{1967Bobeck}. The average over the film thickness $\langle H_{D} \rangle =\frac{1}{h} \int H_{D}(z)dz$ for a cylinder of reversed magnetization relative to the plane (figure 6(c)) simplifies to ~\cite{1971Thiele,1967Bobeck}

\begin{eqnarray}
  \langle H_{D} \rangle   =
\begin{cases}
    \frac{M_{s}h}{R} (1+2 ln \frac{8R}{h}) , \text{at DW for }  R \gg h \\
    4\pi M_{s} \Big(2\sqrt{1+\frac{R^2}{h^2}}-\frac{2R}{h}-1 \Big),  \text{at center}
\end{cases}
\end{eqnarray}

\noindent Then, $p_{D}=2M_{s}\langle H_{D}\rangle$ or a force $F_{D}=4\pi h^{2}M_{s}^{2} (1+2 ln (8R/h))$. The DM field points opposite magnetization at DW (figure 6(c)), will tend to reverse it, and push the DW outwards~\cite{1971Thiele}. The DM field remains comparable to external fields known to influence the AOS dynamics in GdFeCo~\cite{2016Tsema} at relatively large $R/h$. For instance, $H_{D} \approx 0.02\times M_{s}\approx 350~G$ for Co at $R/h = (10~\mu m)/(10~nm)=10^3$.

The DM field of a demagnetized cylinder (figure 6(d)) can be calculated similarly to give

\begin{eqnarray}
  \langle H_{D}' \rangle  =
\begin{cases}
    2 \pi M_s + \frac{h M_{s}}{2R} (1+ 2 ln \frac{8R}{h}) , \text{at DW for }  R \gg h \\
    4\pi M_{s} \Big(\sqrt{1+\frac{R^2}{h^2}}-\frac{R}{h}\Big), \text{at center}
\end{cases}
\end{eqnarray}

\noindent where the approximate expression at the DW can be used with an error $<1\%$ for $R/h>2$. As expected, $\langle H_{D}' \rangle \rightarrow 2 \pi M_s$ or half of the uniform layer $4\pi M_{s}$ at the DW as $R \rightarrow \infty$, and $\langle H_{D}' \rangle \rightarrow 4 \pi M_s$ at the center as $R \rightarrow 0$. The case of a demagnetized cylinder is more complex because exchange and anisotropy energies are different on the DW sides. It is also observed that the demagnetized cylinder quickly evolves into a reversed cylinder when the final $S$ state is obtained (supplementary figure 1).

\begin{figure}
\centering\rotatebox{0}{\includegraphics[scale=0.2]{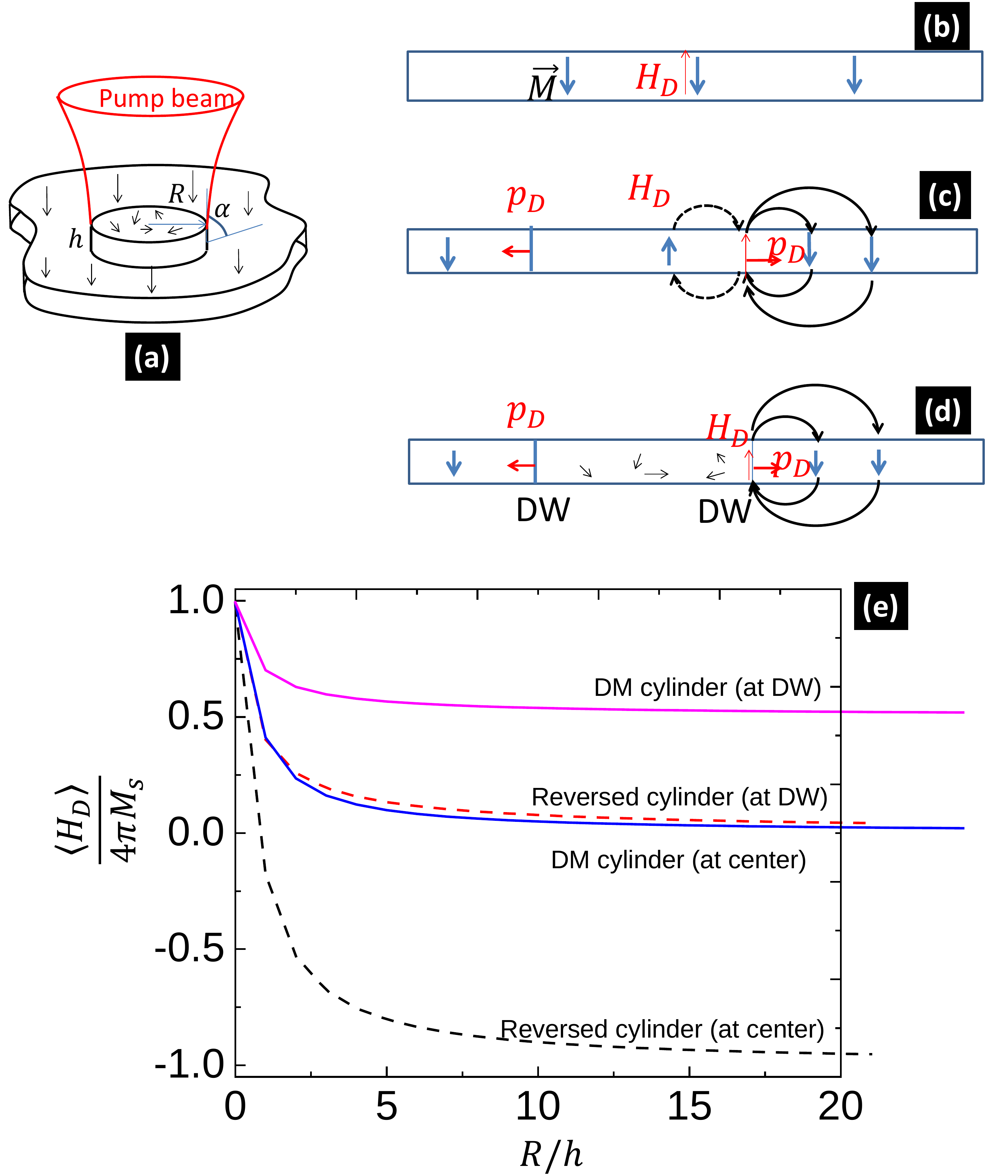}}
\caption{\label{fig:Figure1} (a) Sketch of a demagnetized cylinder of radius $R$ and height $h$ made by the laser beam. (b) Side view of a plane with an uniform PMA magnetic structure with $\langle H_{D}\rangle =4 \pi M_{s}$ inside and no field lines outside. (c) Field lines for a reversed cylinder, where additional (dotted) fields decrease $H_{D}$ compared to the next case. (d) Field lines for a DM cylinder. (e) Dependence on $R/h$ of DM fields at the center and edge of a demagnetized or reversed cylinder. The field at other points is in-between these two limits. Fields of a partially demagnetized cylinder can be obtained by combining these cases.}
\end{figure}

The $p_{w}=p_{D}$ condition for $R=50~nm$, $h=10~nm$ and $M_{s}=1500\times 10^{3}~A/m$ is plotted in figure 5(b). Above it, the inward DW pressure $p_{w}$ is too large and the domain closes. When $M_{s}$ decreases $3$ times, the PMA line decreases $3^{2}$ times and the parabola decreases $3^{4}$ times (figure 5(c)). The boundaries of the $NC$ state calculated with the continuum model are consistent with micromagnetic simulations.

The results show that states to the left of $NC$ states and above the PMA line are switched $S$ or multiple domain $MD$ states, depending on $K$. A random spin state in a cylindrical volume is a sufficient precursor of magnetization reversal within a range of $K$ and $A/a^{2}$ energy densities.

\subsection{Connection to experiments}

The exchange stiffness $A$ is large in practice, and magnetic materials place to the right of the $S/NC$ boundary in figure 5(b-c). For instance, bulk cobalt has $\frac{A}{d^{2}}\approx \frac{21\times 10^{-12}~J/m}{(2.2\times10^{-10}~m)^{2}}=430~\frac{J}{cm^{3}}$ at $T \ll T_{c}$, where $d=2.2~A$ is the Co lattice constant. The micromagnetic simulations confirm that $50~nm$ radius regions with randomly-oriented spins will be closed, or that the PMA ferromagnetic state is stable against these small volume fluctuations, as expected.

Volumes with a radius comparable to the beam diameters are inaccessible in simulations. However, the $MD$ and $NC$ states have been observed in AOS experiments with materials above the PMA condition line. For instance, an $MD$ state~\cite{2014Lambert} is obtained when the equilibrium linear domains are smaller than laser spot size~\cite{2016Hadri-a}. A $NC$ final state has been observed in GdFeCo for beams with $R<5~\mu m$~\cite{2016Gorchon-b}, similar to the case of thicker magnetic bubble materials~\cite{1971Callen}.

This suggests that the $S$ final state of micromagnetic simulations represents the AOS state in these cases. For a narrow range near the $S/NC$ boundary the final state alternated randomly between $S$ and $NC$ in the simulations, similar to the stochastic AOS nucleation in Co/Pt~\cite{2016Medapalli}. AOS domains induced by DM fields appear in polarizing~\cite{2009Ogasawara} and XMCD~\cite{2015Gierster} microscopy. A similar sequence of events to that obtained for a $S$ final state (supplementary material) has been observed in time-resolved images of AOS in GdFeCo~\cite{2014Hashimoto}.

Cumulative AOS experiments are done with multiple pulses, at non-uniform and time-dependent temperatures, which can increase up to $T_{C}$ (section 2). The micromagnetic energy densities $K,A/a^{2},2\pi M_{s}^{2}$ are quantities averaged over spin fluctuations and depend on temperature. Calculations for bulk Co in equilibrium at one temperature give $A\propto [M(T)]^{1.8}$ and $K\propto [M(T)]^{3}$~\cite{2016Moreno}. The temperature dependence of $K, A, M^{2}_{s}$ can be accounted for with additional forces $F_{u,g}$ in a continuum model, for uniform and spatially-dependent $T_{acc}$, respectively. If each pulse gives a DM cylinder and reverses the magnetization the net result of multiple pulses incident on the same area would be difficult to predict and depend on factors outside experimental control. Nevertheless, well-defined reversed domains are observed in cumulative AOS. Micromagnetic simulations confirm that DM fields and energies are reduced as the first reversed domain expands beyond the beam footprint and that successive pulses nucleate domains that collapse, leaving a stable reversed domain.

AOS was not obtained in Co/Pd multilayers when $12~ns$ pulses were used~\cite{2015Stark}. This could be due to a large anisotropy energy, placing the material in the $MD$ region, or to the pulse duration. A lower $1.6~ps$ limit on the pulse duration for AOS in Co/Pt~\cite{2016Medapalli} and an upper $1.5-15~ps$ limit in GdFeCo~\cite{2016Gorchon-b} have been reported. Future work may apply different pulse durations, with micromagnetic simulations of larger volumes, non-uniform and time-dependent $K,A,M_{s}$, accounting for the variation of accumulation temperature $T_{acc}$ and a scanning beam.

\section{Conclusion}

Ultrafast demagnetization was observed in Co/Pd ferromagnetic superlattices. Heat accumulation at high power removes the magnetization. The role of demagnetizing fields in domain nucleation and domain wall motion is quantified with micromagnetic simulations to obtain the conditions for which spins of a small demagnetized cylinder evolve into a reversed domain.

\section*{Acknowledgments}

This research was supported by the University of Louisville Research Foundation. We thank X Wang and J Abersold for assistance with sample preparation at the University of Louisville cleanroom.

\begin{figure}
\centering\rotatebox{0}{\includegraphics[scale=0.6]{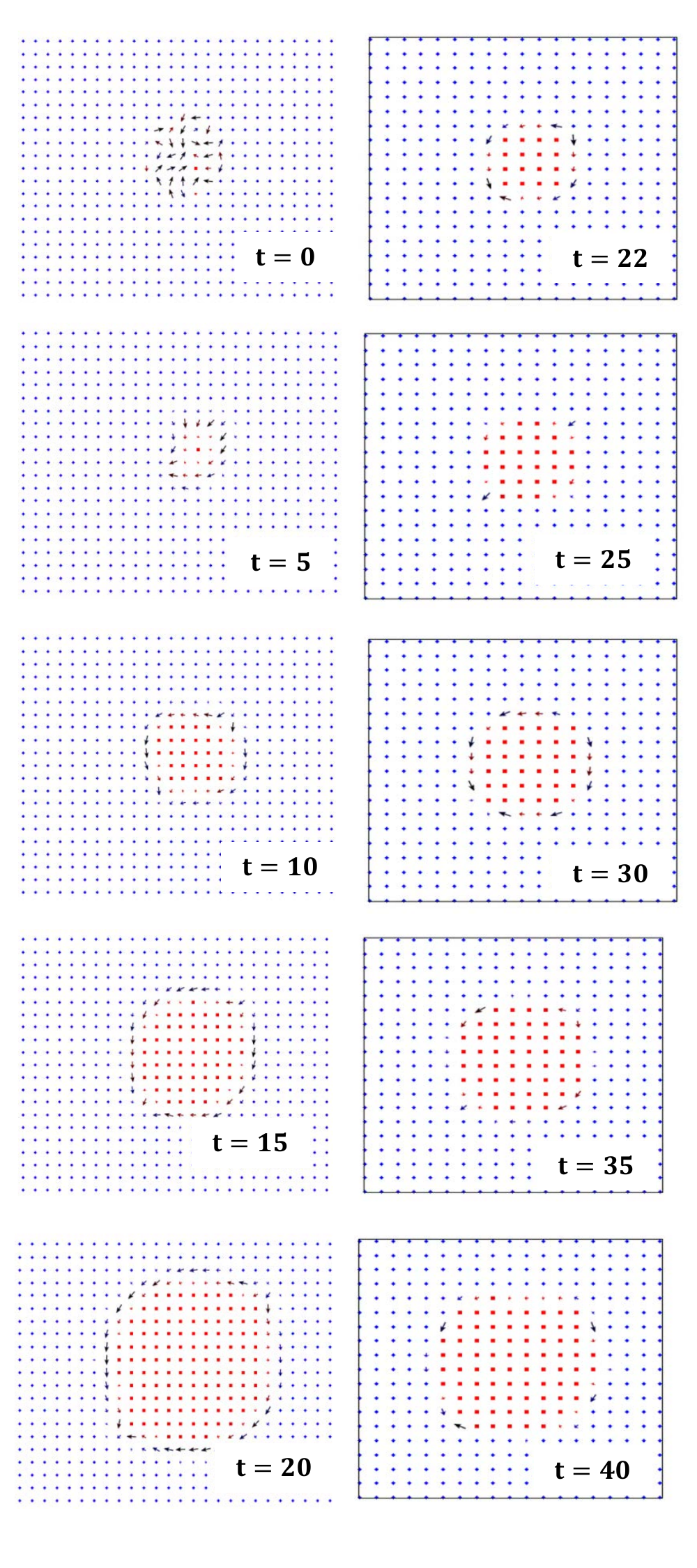}}
\caption{\label{fig:Figure1} \underline{Supplementary figure 1}: Image sequence of the magnetization time-evolution for one $S$ point in figure 5(b) at $K=1800\times10^{3}~J/m^{3}, A/a^{2}=2400\times10^{3}~J/m^{3}, M_{s}=1500\times10^{3}~A/m$. Field of view zooms out at $t=21$ to a larger area.}
\end{figure}

\begin{figure}
\centering\rotatebox{0}{\includegraphics[scale=0.3]{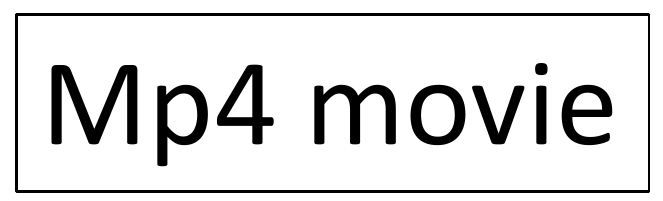}}
\caption{\label{fig:Figure1} \underline{Supplementary material 2}: A movie of the time-evolution from which the image sequence above was taken can be seen at http://www.physics.louisville.edu/smadici/WebpageResearch.html. It will play with VLC software.}
\end{figure}

\end{document}